\documentclass[aps,prb,twocolumn,preprintnumbers,amsmath,amssymb,superscriptaddress]{revtex4-1}

\usepackage{graphicx}
\usepackage{bm}
\usepackage{color}
\usepackage{hyperref}
\usepackage[normalem]{ulem}

\usepackage{t1enc}
\usepackage[polish,english]{babel}
\usepackage[cp1250]{inputenc}
\usepackage{polski}
\usepackage[T1]{fontenc}
\usepackage[normalem]{ulem}

\begin{document}

\title{The Influence of Magnetic Anisotropy on the Kondo Effect and Spin-Polarized Transport
through Magnetic Molecules, Adatoms and Quantum Dots }

\author{Maciej Misiorny}
 \email{misiorny@amu.edu.pl}
\affiliation{Faculty of Physics, Adam Mickiewicz University,
61-614 Pozna\'{n}, Poland}

\author{Ireneusz Weymann}
\affiliation{Faculty of Physics, Adam Mickiewicz University,
61-614 Pozna\'{n}, Poland} \affiliation{Physics Department, Arnold
Sommerfeld Center for Theoretical Physics and Center for
NanoScience, Ludwig-Maximilians-Universit\"{a}t M\"{u}nchen, 80333
M\"{u}nchen, Germany}

\author{J\'{o}zef Barna\'{s}}
\affiliation{Faculty of Physics, Adam
Mickiewicz University, 61-614 Pozna\'{n}, Poland}
\affiliation{Institute of Molecular Physics, Polish Academy of
Sciences, 60-179 Pozna\'{n}, Poland
}%

\date{\today}

\begin{abstract}
Transport properties in the Kondo regime of a nanosystem
displaying uniaxial magnetic anisotropy (such as a magnetic
molecule, magnetic adatom or quantum dot coupled to a localized
magnetic moment) are analyzed theoretically. In particular, the
influence of spin-polarized transport through a local orbital of
the system and exchange coupling of conduction electrons to the
system's magnetic core on the Kondo effect is discussed. The
numerical renormalization group method is applied to calculate the
spectral functions and linear conductance in the case of the
parallel and antiparallel configurations of the electrodes'
magnetic moments. It is shown that both the magnetic anisotropy as
well as the exchange coupling between electrons tunneling through
the conducting orbital and magnetic core play an important role in
formation of the Kondo resonance, leading generally to its
suppression. Specific transport properties of such system appear
also as a nontrivial behavior of tunnel magnetoresistance. It is
also shown that the Kondo effect can be restored by an external
magnetic field in both the parallel and antiparallel magnetic
configurations.
\end{abstract}



\maketitle


\section{Introduction}

Growing trend towards building ever more efficient and smaller
electronic devices inevitably draws the researchers' attention to
nanoscopic hybrid systems. In this respect, single atoms or
molecules seem to be promising prospects, as their incorporation
into electronic nanodevices allows for developing novel systems
capable of performing strictly imprinted
functions,~\cite{Joachim_Nature408/00,Tans_Nature393/98,Park_Nature407/00,Heath_PhysToday56/03,Piva_Nature435/05,Bogani_NatureMater7/08}
among which information storage is of key
interest.~\cite{Green_Nature445/07,Mannini_NatMater8/09,Loth_NatPhysics6/10,Mannini_Nature10}
Consequently, due to recent advances in experimental techniques
enabling to address transport through individual  atoms and
molecules, both natural as well as artificial (quantum dots)
systems exhibiting magnetic anisotropy, such as magnetic atoms of
spin $S>1/2$ (i.e. Fe, Co or
Mn)~\cite{Gambardella_Science300/03,Hirjibehedin_Science317/07,Meier_Science320/08,Otte_NatPhysics4/08,Brune_SurfSci603/09,Loth_NatPhysics6/10}
or single-molecule magnets
(SMMs),~\cite{Gatteschi_book,Heersche_PRL96/06,Jo_NanoLett6/06,Voss_PRB78/08,Zyazin_NanoLett10/10}
have become the object of intensive studies.

It has been suggested, and in the case of magnetic adatoms also
experimentally proven,~\cite{Loth_NatPhysics6/10} that magnetic
state of such systems can by controlled by the use of
spin-polarized
currents~\cite{Elste_PRB73/06,Timm_PRB73/06,Misiorny_PRB75/07,Misiorny_PSS246/09,Misiorny_PRB79/09,FDelgado_PRL104/10}
or spin bias.~\cite{Lu_PRB79/09} This practically means that the
system's magnetic moment can be switched between two metastable
states of minimal energy by only applying an electric/spin current
pulse of a proper amplitude.~\cite{Misiorny_PSS246/09}
Furthermore, if attached to two metallic nonmagnetic electrodes a
SMM can act as a spin
filter.~\cite{Barraza_PRL102/09,Zhu_APL96/10,Hao_APL96/10} If,
however, coupled to electrodes characterized by unequal spin
polarizations, the molecule can reveal transport characteristics
typical of a spin diode.~\cite{Misiorny_EPL89/10} Most of these
results have been obtained in the limit of weak coupling between a
SMM and reservoirs of spin-polarized electrons. Nevertheless, in
some situations, when mixing of localized electron states
responsible for transport properties of the molecule and extended
electron states in electrodes is significant, such an assumption
not necessarily has to be correct.

In the strong coupling regime the electronic correlations can lead
to an additional resonance in the density of states at the Fermi
level of electrodes, known as the \emph{Kondo-Abrikosov-Suhl
resonance}.~\cite{Hewson_book,Kouwenhoven_PhysWorld14/01,Ternes_JPhysCondensMatter21/09}
Since the end of the 1990s, the presence of the Kondo effect has
been successively demonstrated in a large variety of nanoscopic
objects like quantum
dots,~\cite{Goldhaber_Nature391/98,Cronenwett_Science281/98,Sasaki_Nature405/00}
magnetic adatoms,~\cite{Madhavan_Science280/98,Li_PRL80/98}
nanotubes,~\cite{Nygard_Nature408/00} and different types of
molecules: Co(II)-based coordination
complexes;~\cite{Park_Nature417/02} di\-va\-na\-di\-um
molecules;~\cite{Liang_Nature417/02} and C$_{60}$ molecules
attached to gold~\cite{Yu_NanoLett4/04} or ferromagnetic nickel
electrodes.~\cite{Pasupathy_Science306/04} However, in the case of
nanosystems characterized by large spins the prominent role of the
magnetic anisotropy in formation of the Kondo effect has been
experimentally established only very
recently.~\cite{Otte_NatPhysics4/08,Parks_Science328/10} It turned
out that the Kondo effect can be then tuned by changing both the
orientation (e.g. by controlling the adatom's local
environment~\cite{Otte_NatPhysics4/08}) and magnitude (e.g. by
mechanical straining of the molecule~\cite{Parks_Science328/10})
of the magnetic anisotropy. Moreover, Parks \emph{et
al.}~\cite{Parks_Science328/10} were able to tune the anisotropy
constant continuously and to modify accordingly the energy
spectrum underlying the Kondo state. As a result,  they managed to
observe a crossover from the fully screened to underscreened Kondo
effect. It is worthy of note that more recently electric field
control of magnetic anisotropy has been experimentally established
for a SMM embedded into a planar three-terminal
device.~\cite{Zyazin_NanoLett10/10} Although a few theoretical
works focused on transport related issues in SMMs in the Kondo
regime have been already
published,~\cite{Romeike_PRL96I/06,Romeike_PRL97/06,Leuenberger_PRL97/06,Gonzalez_PRB78/08,Roosen_PRL100/08,
Wegewijs_NJPhys9/07,Wang_PRB79/09,Elste_PRB81/10}
experimental evidence of the Kondo effect in transport through
SMMs has been found only very
recently.~\cite{Zyazin_NanoLett10/10}

The earlier works on the Kondo phenomenon in transport through
SMMs have been primarily focused on the role of transversal
magnetic anisotropy, and hence also on the role of quantum
tunneling of the SMM's spin  in the formation of the Kondo state.
It has been shown that the interplay of quantum tunneling and spin
exchange processes between the molecule and tunneling electrons
may result in the pseudo-spin 1/2 Kondo
effect.~\cite{Romeike_PRL96I/06,Romeike_PRL97/06} Furthermore, it
was soon realized that when even a moderate transverse magnetic
field is applied, any qualitative difference  between the
mechanisms  of the Kondo effect for molecules with half- and
full-integer spins cease to exist.~\cite{Leuenberger_PRL97/06}
Mapping of the Anderson-type Hamiltonian describing a SMM onto the
spin-1/2 anisotropic Kondo Hamiltonian~\cite{Gonzalez_PRB78/08}
led to the conclusion that, depending on whether the molecule's
total spin is reduced or augmented in the singly charged state,
the coupling in the Kondo Hamiltonian is antiferromagnetic or
ferromagnetic, respectively. In the former case the Kondo effect
is revealed, whereas in the latter one no resonance at the Fermi
level is present due to renormalization of the transverse coupling
to zero. In addition, the Kondo effect is expected to oscillate as
a function of the magnitude of transverse field due to the
Berry-phase periodical modulation of the tunnel
splitting.~\cite{Leuenberger_PRL97/06} Finally, the nonequilibrium
spin dynamics of a SMM, triggered by a sudden change in the
magnetic field amplitude has been studied, with the main emphasis
on the time evolution of the Kondo
screening.~\cite{Roosen_PRL100/08}

Since physical mechanisms governing Kondo correlations in
spin-polarized transport through nanoscopic systems exhibiting the
magnetic anisotropy, such as magnetic adatoms or SMMs, are still
at the early stage of research, the objective of the present paper
is to provide further insight into the problem. In particular, we
investigate how magnetic anisotropy affects the system's transport
characteristics such as conductance and tunnel magnetoresistance
(TMR) in the linear response regime. To properly describe the
transport properties in the strong  coupling regime, we employ the
Wilson's numerical renormalization group (NRG)
approach.~\cite{Hewson_book,Wilson_RevModPhys47/75,Bulla_RevModPhys80/08}
This method is known as very powerful and essentially exact in
solving quantum impurity problems.

The paper is organized as follows. In Sec. II we describe the
model Hamiltonian used in calculations and provide a brief
introduction to the NRG method. Numerical results and their
discussion are given in Sec. III, where we analyze the spectral
functions of the orbital level as well as the conductance and
tunnel magnetoresistance (TMR) in the linear response regime. The
above quantities are analyzed as functions of the orbital level
position, strength and type of exchange coupling, and the
anisotropy constant. In addition, we also discuss the effect of
external magnetic field. Finally, the summary and conclusions are
given in Sec. IV.

\section{Theoretical description}

\subsection{Model}

\begin{figure}[t]
  \includegraphics[width=0.65\columnwidth]{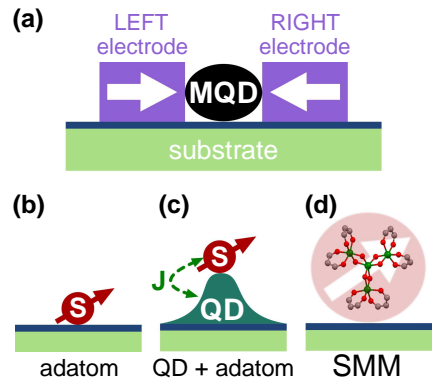}
  \caption{\label{Fig:1} (Color online)
  (a) Schematic representation of the system under consideration.
  The system consist of two ferromagnetic electrodes to which
  a magnetic quantum dot (MQD) exhibiting magnetic anisotropy is attached.
  As the MQD one can conceive either a magnetic adatom (i.e. Fe, Co, Mn) (b), semiconductor quantum dot
  coupled to a magnetic moment (c), or a single-molecule magnet (SMM) (d) -- here,
  only the magnetic core of the Fe$_4$ molecule~\cite{Mannini_AdvMater21/09} is schematically depicted.}
\end{figure}

We consider a generic theoretical model that allows for capturing
essential features of quantum objects such as magnetic adatoms,
quantum dots coupled to localized magnetic moments, and SMMs, see
Fig.~\ref{Fig:1}. It is assumed that electronic transport takes
place \emph{via} a single local orbital level (OL) of the system
(conducting orbital of a SMM, adatom or quantum dot), which is
coupled to electrodes. Moreover, the OL is also exchange-coupled
to the corresponding magnetic core. Without loss of generality, we
will henceforth refer to the systems under investigation as
\emph{magnetic quantum dots} (MQDs).

The total Hamiltonian of a MQD coupled to external leads can be
written as
    \begin{equation}\label{Eq:1}
    \mathcal{H} = \mathcal{H}_\textrm{MQD} + \mathcal{H}_\textrm{leads} + \mathcal{H}_\textrm{tun}.
    \end{equation}
The first term represents the MQD and
has the form~\cite{Elste_PRB73/06,Timm_PRB73/06,Misiorny_PRB75/07,Misiorny_PSS246/09,Misiorny_PRB79/09}
    \begin{align}\label{Eq:2}
    \mathcal{H}_\textrm{MQD} =& -DS_z^2
    + \sum_{\sigma=\uparrow,\downarrow} \varepsilon\,n_\sigma  + U\, n_\uparrow n_\downarrow
    \nonumber\\
    &- J \textbf{s}\cdot\textbf{S} + B_z(S_z+s_z),
    \end{align}
where $D$ stands for the uniaxial anisotropy constant of the MQD,
while $S_z$ denotes the $z$th component of the MQD's internal spin
operator $\bm{S}$. Since in the present paper we focus only on
systems displaying magnetic bistability, the anisotropy constant
is assumed to be positive ($D>0$). Furthermore,
$n_\sigma=c_\sigma^\dagger c_\sigma^{}$ is the OL occupation
operator, where $c_\sigma^\dagger (c_\sigma^{})$ creates
(annihilates) a spin-$\sigma$ electron of energy $\varepsilon$ in the OL. The
Coulomb energy of two electrons of opposite spins occupying the OL
is given by $U$. The penultimate term of Eq.~(\ref{Eq:2}) accounts
for exchange coupling between the magnetic core of a MQD and the
spin of an electron in the OL, represented by
$\textbf{s}=\frac{1}{2}\sum_{\sigma\sigma'}c_\sigma^\dag
\bm{\sigma}_{\sigma\sigma'}^{} c_{\sigma'}^{}$, where
$\bm{\sigma}\equiv(\sigma^x,\sigma^y,\sigma^z)$ is the Pauli spin
operator. The $J$-coupling can be either of \emph{ferromagnetic}
($J>0$) or \emph{antiferromagnetic} ($J<0$) type.  Finally, the
last term of Eq.~(\ref{Eq:2}) describes the Zeeman interaction of the MQD
with an external magnetic field $\bm{B}=(0,0,B_z)$ oriented along
the easy axis of a MQD. Note that we  put here
$g\mu_\textrm{B}\equiv1$.

The ferromagnetic metallic electrodes, to which a MQD is coupled
through the OL, are characterized by noninteracting itinerant
electrons with the dispersion relation
$\varepsilon_{\bm{k}\sigma}^q$, where $q$ indicates either left
($q=L$) or right ($q=R$) electrode, $\bm{k}$ denotes a wave vector
and $\sigma$ is a spin index of an electron. Thus, the leads'
Hamiltonian is given by
    \begin{equation}\label{Eq:3}
    \mathcal{H}_{\textrm{leads}}=\sum_{q=L,R}\sum_{\bm{k}}\sum_{\sigma=\uparrow,\downarrow}\varepsilon_{\bm{k}\sigma}^q
    a_{\bm{k}\sigma}^{q\dag} a_{\bm{k}\sigma}^q,
    \end{equation}
with  $a_{\bm{k}\sigma}^{q\dag}$ ($a_{\bm{k}\sigma}^q$) being the
relevant electron creation (annihilation) operator. At this point,
it should also be mentioned that in the present paper we limit the
discussion to collinear (parallel and antiparallel) configurations
of electrodes' magnetic moments. Furthermore, the MQD's easy axis is
assumed to be collinear with these moments as well. Finally,
electron tunneling processes between the MQD and electrodes are
included in the term $\mathcal{H}_\textrm{tun}$,
    \begin{equation}\label{Eq:4}
    \mathcal{H}_\textrm{tun} = \sum_{q=L,R}\sum_{\bm{k}}\sum_{\sigma=\uparrow,\downarrow}T_{\bm{k}\sigma}^q a_{\bm{k}\sigma}^{q\dag} c_\sigma^{} + \textrm{H.c.}
    \end{equation}
where $T_{\bm{k}\sigma}^q$ denotes the tunnel matrix element
between the OL and the $q$th lead.

In the linear response regime, it is numerically convenient to
introduce the following canonical
transformation,~\cite{Glazman_JETPLett47/88,Ng_PRL61/88,Bruus_book}
    \begin{equation}\label{Eq:5}
    \left(
        \begin{array}{c}
        a_{\bm{k}\sigma}^e\\
        a_{\bm{k}\sigma}^o
        \end{array}
    \right)
    =\frac{1}{\mathcal{V}_{\bm{k}\sigma}}
    \left(
        \begin{array}{cc}
        T_{\bm{k}\sigma}^{L} & T_{\bm{k}\sigma}^{R} \\
        -T_{\bm{k}\sigma}^{R} & T_{\bm{k}\sigma}^{L}
        \end{array}
    \right)
    \left(
        \begin{array}{c}
        a_{\bm{k}\sigma}^L\\
        a_{\bm{k}\sigma}^R
        \end{array}
    \right),
    \end{equation}
where,
$\mathcal{V}_{\bm{k}\sigma}=\sqrt{|T_{\bm{k}\sigma}^L|^2+|T_{\bm{k}\sigma}^R|^2}$,
and the label $e\ (o)$ denotes the even (odd) combination of the
leads operators. Such a rotation in the space of the left-right
electron operators results in separation of the total Hamiltonian,
Eq.~(\ref{Eq:1}), into two independent parts. The first one
involves the OL coupled to a single electron reservoir described
by the even linear combination of the leads' electron operators,
$a_{\bm{k}\sigma}^e$, while the other one is related with
non-interacting electrons decoupled from MQD and described by the
odd operators $a_{\bm{k}\sigma}^o$. Consequently, the tunneling
Hamiltonian can be written as
    \begin{equation}\label{Eq:6}
    \mathcal{H}_\textrm{tun}=\sum_{\bm{k}}\sum_{\sigma=\uparrow,\downarrow}\mathcal{V}_{\bm{k}\sigma}\left[a_{\bm{k}\sigma}^{e\dag}c_{\sigma}^{}
    +c_{\sigma}^{\dag}a_{\bm{k}\sigma}^{e}\right],
    \end{equation}
with $\mathcal{V}_{\bm{k}\sigma}$ being effective OL-lead
tunneling matrix elements. In the following we assume that the
full spin-dependence is taken into account through the matrix
elements
$\mathcal{V}_{\bm{k}\sigma}$.~\cite{Choi_PRL92/04,Sindel_PRB76/07}
For simplicity, we assume a flat conduction band in the interval
$[-W,W]$, so that $\rho_\sigma(\omega)
\equiv \rho = \frac{1}{2W}$, with $W$ representing the cut-off energy of the system and $W\equiv1$
taken as the energy unit. Finally, the energy dependence of
$\mathcal{V}_{\bm{k}\sigma}$ is neglected,
$\mathcal{V}_{\bm{k}\sigma}\equiv\mathcal{V}_{\sigma}$.\cite{Krishna_PRB21I/80}
Under these circumstances, the overall effect of the ferromagnetic
reservoir on the MQD is completely determined by the hybridization
function $\Gamma_\sigma$,
\begin{equation}\label{Eq:7}
    \Gamma_\sigma=\pi\rho|\mathcal{V}_{\sigma}|^2.
\end{equation}

\subsection{Method of calculations}

In order to determine transport properties in the strong coupling
regime, we use the Wilson's numerical renormalization group
method.~\cite{Hewson_book,Wilson_RevModPhys47/75,Bulla_RevModPhys80/08}
The NRG technique consists of logarithmic discretization of the
conduction band (with a discretization parameter $\Lambda>1$) into
intervals
$[\Lambda^{-(n+1)}W,\Lambda^{-n}W]$ and
$[-\Lambda^{-n}W,-\Lambda^{-(n+1)}W]$ for
$n=0,1,2,3,\ldots$, which allows for resolving transport
properties on energy scales logarithmically approaching the Fermi
level. After having discretized the conduction band, such a model
is mapped onto a semi-infinite chain, whose first site is coupled to the impurity (in our
case the MQD). The Hamiltonian
then reads~\cite{Hewson_book,Bulla_RevModPhys80/08}
    \begin{align}\label{Eq:8}
    \mathcal{H}=\mathcal{H}_\textrm{MQD} &+ \sum_{\sigma=\uparrow,\downarrow} \sqrt{\frac{\Gamma_\sigma}{\pi\rho}}\big[c_\sigma^\dag f_{0\sigma}^{}+f_{0\sigma}^\dag c_\sigma^{}\big]
    \nonumber\\
    &+\sum_{n=0}^{\infty} \sum_{\sigma=\uparrow,\downarrow}  t_n\big[f_{n\sigma}^\dag f_{n+1\sigma}^{}
    +f_{n+1\sigma}^\dag f_{n\sigma}^{}\big].
    \end{align}
The operators $f_{n\sigma}^{}$ ($f_{n\sigma}^{\dag}$) correspond
to the $n$th site of the Wilson chain, with exponentially decaying
hopping matrix elements $t_n$ between neighboring sites of the
chain.~\cite{Bulla_RevModPhys80/08} As a consequence, by adding
consecutive sites, one is able to access transport at lower and
lower energy scales. In this way the method generally provides a
non-perturbative description of the crossover from a free magnetic
impurity at high temperatures to a screened spin at low
temperatures.~\cite{Bulla_RevModPhys80/08} The Hamiltonian
~(\ref{Eq:8}) can be solved iteratively by adding consecutive
sites of the chain. This procedure allows for resolving static and
dynamic properties of the system at energy scale $\Lambda^{-n/2}$,
with $n$ being a given iteration.

Since the NRG calculations may in general pose a serious numerical
challenge, it becomes essential to take advantage of as many
available symmetries of the system's Hamiltonian as possible.
To efficiently address the present problem, we have employed the
flexible density-matrix numerical renormalization group (DM-NRG)
code,~\cite{Legeza_DMNRGmanual} which can exploit an arbitrary
number of both Abelian and non-Abelian
symmetries.~\cite{Toth_PRB78/08}  In the case under discussion,
the $U_\textrm{charge}(1)\times U_\textrm{spin}(1)$ symmetry of
the model has been used, so that the $z$th component of the total
spin,
\begin{equation}
\widetilde{S}_z^t = S_z^t + \frac{1}{2}\sum_{n=0}^\infty\left( f_{n\uparrow}^\dag f_{n\uparrow}^{} - f_{n\downarrow}^\dag f_{n\downarrow}^{} \right) \,,
\end{equation}
where $S_z^t=S_z+s_z$, as well as the total charge
\begin{equation}
\widetilde{Q}^t = \sum_\sigma c_\sigma^\dag c_\sigma^{} -1 + \sum_{n=0}^\infty\left(\sum_\sigma f_{n\sigma}^\dag f_{n\sigma}^{} -1 \right) \,,
\end{equation}
served as quantum numbers according to which the states of the
Hamiltonian were classified during computation. Finally, the
discretization parameter $\Lambda=1.8$ has been taken in
calculations, and we have kept 2000 states after each step of the
iteration.

The central quantity of interest  is the OL spin-dependent
\emph{spectral
function}~\cite{Bulla_RevModPhys80/08,Costi_JPhysCondensMatter6/94}
\begin{equation}\label{Eq:12}
    A_\sigma(\omega)=-\frac{1}{\pi}\textrm{Im}\,\langle\langle c_{\sigma}^{} | c_\sigma^\dag\rangle\rangle_\omega^\textrm{r}\,,
\end{equation}
where $\langle\langle c_{\sigma}^{} |
c_\sigma^\dag\rangle\rangle_\omega^\textrm{r}$ denotes the Fourier
transform of the retarded Green's function $\langle\langle
c_{\sigma}^{} |
c_\sigma^\dag\rangle\rangle_t^\textrm{r}=-i\theta(t)\big\langle\{c_\sigma^{}(t),c_\sigma^\dag(0)\}\big\rangle$
of the orbital level. Some technical details concerning
calculation of the spectral function can be found in
Appendix~\ref{APP:A}.

Having found the spectral function, one can determine the
\emph{linear response conductance} $g$ from the
Landauer-Wingreen-Meir
formula,~\cite{Landauer_PhilosMag21/70,Meir_PRL66/91,Meir_PRL68/92,Jauho_book}
which at $T=0$ yields
\begin{equation}\label{Eq:13}
    g=\pi\sum_\sigma \frac{2\Gamma_\sigma^L\Gamma_\sigma^R}{\Gamma_\sigma^L+\Gamma_\sigma^R}\cdot A_\sigma(\omega=0)\ \ \ (\textrm{in units of}\ \tfrac{2e^2}{h}),
\end{equation}
with $\Gamma_{\uparrow(\downarrow)}^L=\tfrac{\Gamma}{2}(1\pm P)$
and $\Gamma_{\uparrow(\downarrow)}^R=\tfrac{\Gamma}{2}(1\pm P)$
for the \emph{parallel} magnetic configuration of electrodes,
while $\Gamma_{\uparrow(\downarrow)}^L=\tfrac{\Gamma}{2}(1\pm P)$
and $\Gamma_{\uparrow(\downarrow)}^R=\tfrac{\Gamma}{2}(1\mp P)$
for the \emph{antiparallel} one. We assumed that both electrodes
are made of the same material and $P$ denotes their spin
polarization. The effective coupling between the MQD and the
reservoir for the \emph{parallel} magnetic configuration is
$\Gamma_{\uparrow(\downarrow)}^\textrm{P}=\Gamma(1\pm P)$, while
for the \emph{antiparallel} one
$\Gamma_{\uparrow(\downarrow)}^\textrm{AP}=\Gamma$, with
$\Gamma=(\Gamma_\uparrow+\Gamma_\downarrow)/2$. Note that in the
case of left-right symmetric systems,
$\Gamma_{\uparrow(\downarrow)}^\textrm{AP}$ is independent of
$\sigma$ in the antiparallel configuration, and hence the system
behaves effectively as coupled to nonmagnetic leads. As a result,
the linear spin-resolved conductance (in units of
$\tfrac{2e^2}{h}$) in the two magnetic configurations is given by
\begin{equation}\label{Eq:14}
    \left\{
    \begin{aligned}
        g_{\uparrow(\downarrow)}^\textrm{P}&=\frac{\pi}{2}(1\pm P)\Gamma A_{\uparrow(\downarrow)}^\textrm{P}(\omega=0),\\
        g_{\uparrow(\downarrow)}^\textrm{AP}&=\frac{\pi}{2}(1-P^2)\Gamma A_{\uparrow(\downarrow)}^\textrm{AP}(\omega=0),
    \end{aligned}
    \right.
\end{equation}
with $A_{\sigma}^\textrm{P/AP}$ being the spectral function in the respective magnetic configuration.

\section{Numerical results and discussion}

Transport characteristics of the system, such as the linear-response conductance $g$ and tunnel magnetoresistance
(TMR), have been numerically calculated for a hypothetical MQD
characterized by the spin number $S=2$. Since the key feature of
the transport regime under discussion is the presence of the Kondo
resonance, it is convenient to introduce the Kondo temperature
$T_\textrm{K}$ as the most relevant energy scale for the system,
to which other energy parameters will be related, if necessary. In
the present work, the Kondo temperature $T_\textrm{K}$ is
estimated from the half-width at half-maximum of the Kondo
resonance in spectral function at $T=0$ for $J=0$ and
$P=0$.~\cite{Hewson_book,Ternes_JPhysCondensMatter21/09} We note
here that, for the sake of simplicity, we assume
$k_\textrm{B}\equiv1$, i.e. temperatures are also given in units
of energy. As a result, for the parameres used in
Fig.~\ref{Fig:2}, we get $T_\textrm{K}/W\approx 5\cdot10^{-4}$ ($T_\textrm{K}/\Gamma\approx 0.022$).

Before presenting and discussing numerical results, it is worth recalling that the
Kondo effect appears as a result of spin
exchange processes in the OL due to its strong coupling to
electrodes. This coupling is described here by the effective
hybridization parameter $\Gamma$, which introduces the Kondo temperature
$T_\textrm{K}$ as the relevant energy scale. However, the model considered
provides also another independent physical mechanism (channel)
through which the Kondo effect can be modified, i.e. the exchange
interaction $J$ between an electron in the OL and the MQD's
magnetic core. Therefore, one can expect the ratio of the
$J$-coupling and the Kondo temperature $T_\textrm{K}$ to be the key
parameter controlling whether the Kondo resonance will appear or
not. Indeed, we show below that the electronic correlations
between the OL and electrodes effectively result in formation of
the Kondo resonance only if $|J|\lesssim T_\textrm{K}$.
In order to see how the parameters of the system, especially
the $J$-coupling and the magnetic anisotropy $D$, influence the
transport properties, we first calculate and discuss the relevant
spectral functions.

\subsection{Spectral functions}

\begin{figure*}[t]
  \includegraphics[width=1.75\columnwidth]{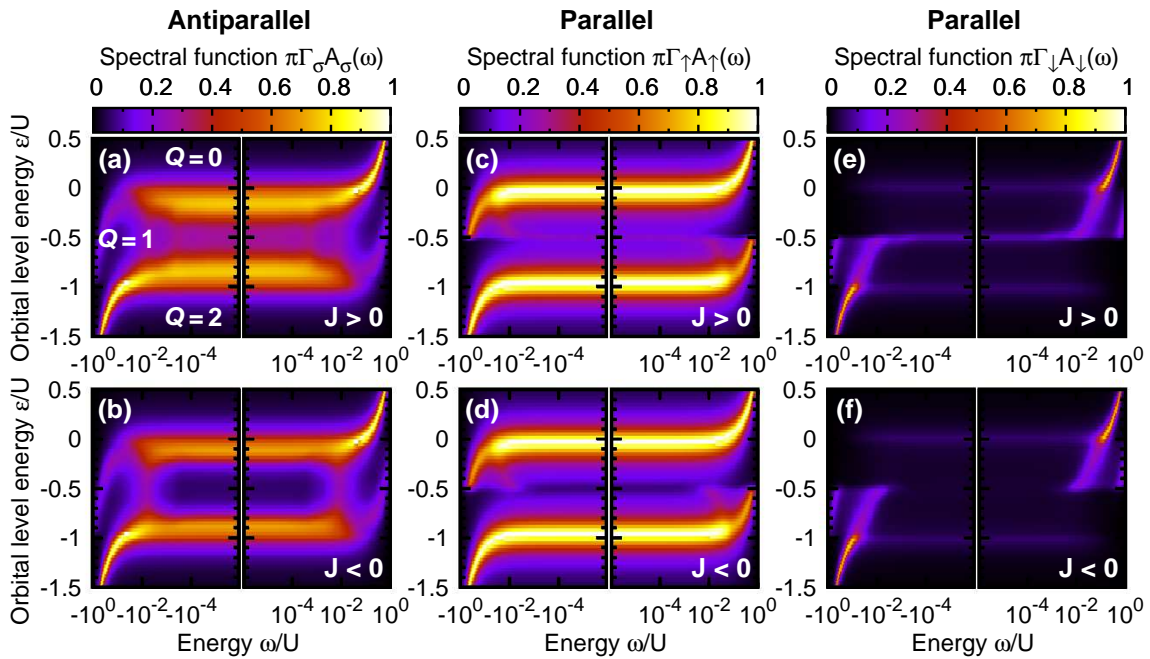}
  \caption{\label{Fig:2} (Color online) Normalized spin-resolved orbital
level (OL) spectral functions, $\pi\Gamma_\sigma
A_\sigma(\omega)$, shown as a function of the OL energy
$\varepsilon$ in the \emph{antiparallel} (a)-(b) and
\emph{parallel} (c)-(f) magnetic configurations. Top panel
corresponds to the case of \emph{ferromagnetic} ($J>0$) coupling
between electrons in the OL and  MQD's magnetic core, while the
bottom one presents results for the \emph{antiferromagnetic}
coupling. The variable $Q$ denotes the average number of electrons
occupying the OL. The parameters are: $U/W=0.3$, $D/U\approx1.7\cdot10^{-4}$ ($D/T_\textrm{K}=0.1$), $\Gamma/U\approx0.075$,
$|J|/\Gamma\approx0.044$ ($|J|/T_\textrm{K}\approx2$), $B_z=0$, and $P=0.5$.
Note that the spectral functions are presented in a logarithmic scale.
  }
\end{figure*}

In the following we will analyze dependence of the spectral
functions of the OL on some essential parameters of the system.
Figure~\ref{Fig:2} shows the normalized spin-resolved spectral
functions $\pi\Gamma_\sigma A_\sigma(\omega)$ as a function of the
OL energy $\varepsilon$  in the antiparallel and parallel magnetic
configurations for $|J|/T_\textrm{K}\approx2$. In the antiparallel
configuration the spectral functions for both spin components are
equal. In the parallel configuration, on the other hand, the main
contribution comes from the spin-up electrons which are the
majority ones. In both configurations and for both types of
exchange coupling, the spectral functions show clear resonances
associated with degeneracy of the neighboring charge states --
compare the boundaries between regions corresponding to different
$Q$, where $Q$ denotes the average number of electrons occupying
the OL. In the singly-occupied regime, $Q=1$, the Kondo effect due
to hybridization of the OL spin with the conduction electrons of
the leads should be present. However, there are two ingredients
that may generally suppress the Kondo resonance: the exchange
coupling $J$ and the exchange field due to the presence of
ferromagnetic leads. Since the results in Fig.~\ref{Fig:2} are
shown for $|J|/T_\textrm{K}\approx2$, only the reminiscent of the
Kondo effect can be observed. The suppression of the Kondo
resonance is especially visible for the antiferromagnetic coupling
($J<0$). Moreover, while in the case of ferromagnetic coupling
($J>0$) some residual Kondo resonance can still be visible for the
energy corresponding to the particle-hole symmetry
($\varepsilon=-U/2$), the resonance is practically absent for the
antiferromagnetic coupling, though some side resonances appear.
The origin of these additional features will be discussed further
in the text.

\begin{figure}[t]
\includegraphics[width=0.99\columnwidth]{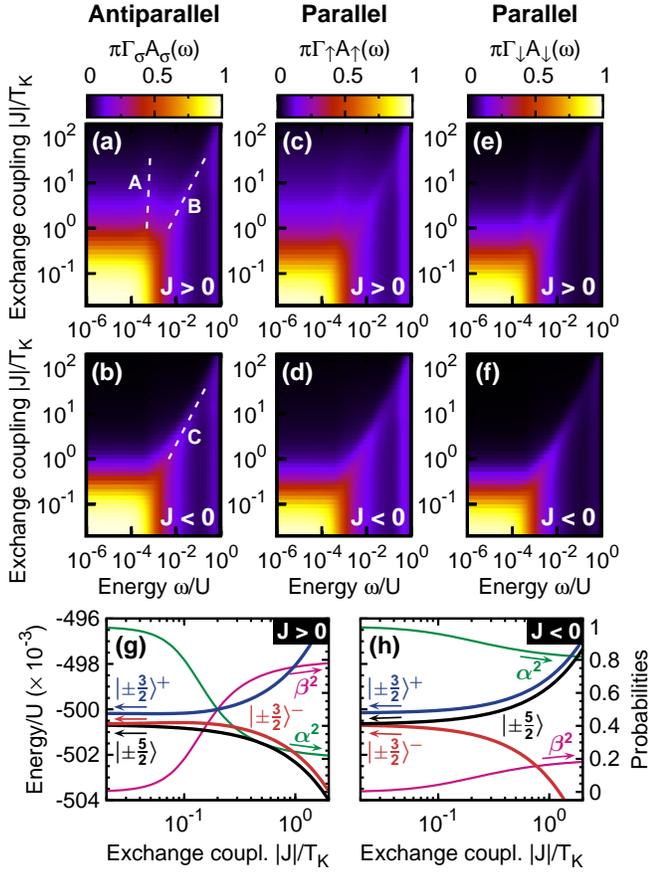}
\caption{\label{Fig:3} (Color online) Normalized spin-resolved
orbital level (OL) spectral functions shown for different values
of the exchange parameter $J$ in the case of (a,c,e)
\emph{ferromagnetic} ($J>0$) and (b,d,f) \emph{antiferromagnetic}
($J<0$) coupling, and for $\varepsilon=-U/2$.
The spectral function displays two additional resonances for $J>0$,
marked by dashed lines A and B, and a single resonance for $J<0$,
marked by a dashed line C.
The bottom panel (g,h) illustrates the dependence of several lowest energy
states of a singly occupied MQD on the parameter $J$ (thick
lines), and the corresponding probabilities of finding the
electron in a certain spin-state for
$|\!\pm\tfrac{3}{2}\rangle^{\pm}$ (thin lines):
$\alpha^2\equiv(\mathbb{A}^+)^2=(\mathbb{B}^-)^2$, while
$\beta^2\equiv(\mathbb{B}^+)^2=(\mathbb{A}^-)^2$ [for details see
Eq.~(\ref{Eq:15}) and the paragraph below it]. Remaining
parameters as in Fig.~\ref{Fig:2}.}
\end{figure}

Let us now analyze how the shape of spectral functions in the
Kondo regime evolves when the exchange interaction between
electrons in the OL and the MQD's core is turned on gradually, see
Figs.~\ref{Fig:3}(a)-(f). Note that whenever we consider behavior
of the system in the particle-hole symmetric point, i.e. for
$\varepsilon=-U/2$, only the range of positive energies is
presented. As one might expect, the behavior of the system for
small values of $|J|$, i.e. $|J|\ll T_\textrm{K}$, resembles that of a
single-level quantum dot, and a well defined and pronounced Kondo
peak in the \emph{antiparallel} configuration of the electrodes'
magnetic moments is observed for $\omega\lesssim T_\textrm{K}$,
see Figs.~\ref{Fig:3}(a)-(b). In the \emph{parallel}
configuration, on the other hand, spin-dependent coupling to the
electrodes acts as an effective exchange
field~\cite{Martinek_PRL91/03} leading to spin splitting of the
OL. This in turn results in suppression of the Kondo effect,
except for the particle-hole symmetric point, $\varepsilon=-U/2$,
as shown in Figs.~\ref{Fig:3}(c)-(f), where the effective exchange
field vanishes. In the antiparallel configuration, the resultant
coupling is the same for the spin up and spin down when the system
is left-right symmetric. Consequently, there is no exchange field
and we observe a well-pronounced peak in the spectral function at
the Fermi level also outside the particle-hole symmetric
point.~\cite{Martinek_PRL91/03,Swirkowicz_JPCM18/06,Swirkowicz_PRB73/06,Barnas_JPCM20/08,Weymann_PRB2011}

The height of the Kondo peak becomes reduced with the increase of
$|J|$, and for $|J|\gg T_\textrm{K}$ the peak almost completely
vanishes. We note that the disappearance of the resonance is
faster in the case of the antiferromagnetic coupling ($J<0$), see
Figs.~\ref{Fig:3}(b,d,f). Furthermore, as the $J$-coupling grows,
some additional features in the spectral functions emerge. Apart
from the Hubbard peak originating from the Coulomb repulsion of
two electrons in the OL level, there are two additional resonances
for $J>0$, marked as dashed lines A and B in Fig.~\ref{Fig:3}(a),
and one resonance for $J<0$, line C in Fig.~\ref{Fig:3}(b).
Interestingly,  position of one of the  two resonances for $J>0$
remains roughly independent of energy (line A), whereas the other
resonance moves towards larger energies as $J$ increases (line B),
see Fig.~\ref{Fig:3}(a).

Some insight into physical origin of the resonances A, B and C can
be gained by considering the lowest energy states of a
free-standing MQD with one extra electron in the OL, see
Figs.~\ref{Fig:3}(g,h).~\footnote{Analytical expressions
describing energies and corresponding states of the systems
represented by the Hamiltonian~(\ref{Eq:2}) can be found in
Refs.~\onlinecite{Timm_PRB73/06}
and~\onlinecite{Misiorny_PSS246/09}. Note, however, that the
notation used in the present paper differs from the notation
employed in the aforementioned articles.} First, we note that the
consequence of exchange interaction between an electron in OL and
magnetic core is a decomposition of the molecular magnetic states
into two spin-multiplets, corresponding to $S+1/2$ and $S-1/2$. In
addition, the sign of the coupling parameter $J$ determines which
of the two multiplets has lower energy. Since  we focus
exclusively on the case of $T=0$, it is justified to take into
account only the relevant low energy states in both
spin-multiplets. These are presented in Figs.~\ref{Fig:3}(g,h),
where the zero-field ($B_z=0$) energy of states
$|\!\pm\frac{5}{2}\rangle$, and $|\!\pm\frac{3}{2}\rangle^\pm$ is
presented as a function of the coupling parameter $J$. The
superscript $\pm$ at the states $|\!\pm\frac{3}{2}\rangle^\pm$ is
used to distinguish between states of higher ($+$) and lower ($-$)
energy. We note that the state $|\!\pm\frac{5}{2}\rangle$ belongs
to the spin-multiplet corresponding to  $S+1/2$, while for $J>0$
the state $|\!\pm\frac{3}{2}\rangle^-$ belongs to the multiplet
$S+1/2$ and the state $|\!\pm\frac{3}{2}\rangle^+$ to the
multiplet $S-1/2$ (note we assumed $S=2$). For $J<0$, the situation
is opposite, i.e. $|\!\pm\frac{3}{2}\rangle^-$ belongs to the
multiplet $S-1/2$ whereas the state $|\!\pm\frac{3}{2}\rangle^+$
to the multiplet $S+1/2$. In the absence of external magnetic
field the  MQD's states of interest take the form
\begin{equation}\label{Eq:15}
    \left\{
    \begin{aligned}
        &\Big|S_z^t=\pm\tfrac{5}{2}\Big\rangle=|\!\uparrow\!(\downarrow)\rangle_\textrm{OL}\otimes|\!\pm2\rangle_\textrm{core},\\
        &\Big|S_z^t=-\tfrac{3}{2}\Big\rangle^{\pm}=
        \mathbb{A}_{-3/2}^{\pm}|\!\downarrow\rangle_\textrm{OL}\otimes|\!-1\rangle_\textrm{core}\\
        &\hspace*{56pt}+\mathbb{B}_{-3/2}^{\pm}|\!\uparrow\rangle_\textrm{OL}\otimes|\!-2\rangle_\textrm{core},
        \\
        &\Big|S_z^t=+\tfrac{3}{2}\Big\rangle^{\pm}=
        \mathbb{A}_{+3/2}^{\pm}|\!\uparrow\rangle_\textrm{OL}\otimes|\!+1\rangle_\textrm{core}\\
        &\hspace*{56pt}+\mathbb{B}_{+3/2}^{\pm}|\!\downarrow\rangle_\textrm{OL}\otimes|\!+2\rangle_\textrm{core},
    \end{aligned}
    \right.
\end{equation}
with $|\bullet\rangle_{\textrm{OL(core)}}$ denoting the spin state
of OL (magnetic core). The coefficients
$\mathbb{A}_m^\pm=\mathbb{A}^\pm\exp[i\phi_m^\pm]$ and
$\mathbb{B}_m^\pm=\mathbb{B}^\pm\exp[i\gamma_m^\pm]$ can be
regarded as effective Clebsch-Gordan coefficients, which are
nontrivial functions of the system's parameters $J$ and
$D$.~\cite{Note1} Here, $\phi_m^\pm$ and $\gamma_m^\pm$ are
relevant phase factors. As above, the superscript $\pm$ is used to
distinguish between states of higher ($+$) and lower ($-$) energy.
It is worth emphasizing that, to some extent, the situation under
consideration is similar to the case of a quantum dot subjected to
an external magnetic field which leads to splitting of the Kondo
resonance.~\cite{Goldhaber_Nature391/98,Martinek_PRL91/03}
However, in the case of a simple quantum dot in a magnetic field
there are two energy levels, whereas the energy structure of a MQD
is much more complex, even in zero field. Moreover, except for the
states $|\!\pm\frac{5}{2}\rangle$, all MQD's states for $Q=1$
correspond to an electron in the OL being in the superposition of
spin-up and spin-down states.

Let us now discuss cotunneling processes leading to the resonances
for  \emph{ferromagnetic} (FM) exchange coupling ($J>0$), the top panel
of Fig.~\ref{Fig:3}. We emphasize that now we consider the regime
of large $J$ (i.e. $J>T_\textrm{K}$ and $J<\Gamma$), where the zero-energy resonance is absent and only
side resonances (lines A and B) appear. Assume that initially the
molecule occupies the state $|\!-\frac{5}{2}\rangle$. Due to
spin-flip cotunneling processes, the MQD can be  excited to one of
the two states: $|\!-\frac{3}{2}\rangle^-$ (resonance A) and
$|\!-\frac{3}{2}\rangle^+$ (resonance B), see Fig.~\ref{Fig:3}(g).
Taking into account the energy spectrum~\cite{Note1} one can
estimate the corresponding energy gaps for $|J|\gg D$ as (exact
formulas in Appendix~\ref{APP:B})
\begin{equation}\label{Eq:16}
    \left\{
        \begin{aligned}
        \Delta_1^\textrm{FM}&\approx 2SD\Bigg[1-\frac{2(|J|-D)}{(2S+1)(|J|-2D)}\Bigg],
        \\
        \Delta_2^\textrm{FM}&\approx\frac{2S+1}{2}|J|.
        \end{aligned}
    \right.
\end{equation}
Thus, the resonance corresponding to the line A in
Fig.~\ref{Fig:3}(a) is related to transitions characterized by the
energy gap $\Delta_1^\textrm{FM}$, while the resonance indicated by
the line B is associated with the gap $\Delta_2^\textrm{FM}$.
Moreover, one can note that $\Delta_2^\textrm{FM}$ depends linearly
on $J$, whereas $\Delta_1^\textrm{FM}$ only slightly changes with
$J$.

The picture presented above for $J>0$ alters only slightly when
the exchange coupling changes to the \emph{antiferromagnetic} (AFM) one
($J<0$), see Figs.~\ref{Fig:3}(b,d,f). Position of the
spin-multiplets $S+1/2$ and $S-1/2$ is now interchanged when
compared to that for $J>0$, Fig.~\ref{Fig:3}(h). When the MQD is
initially in the state $|\!-\frac{3}{2}\rangle^-$, then the
spin-flip cotunneling processes can excite the MQD to one of the
states: $|\!-\frac{5}{2}\rangle$ and $|\!-\frac{3}{2}\rangle^+$.
Moreover, for $|J|\gg D$, the energy gaps associated with these
transitions are roughly equal,
\begin{equation}\label{Eq:17}
        \Delta_1^\textrm{AFM}\approx \Delta_2^\textrm{AFM}\approx\frac{2S+1}{2}|J|.
\end{equation}
As a consequence, only the resonance denoted by the line C is then
visible in Fig.~\ref{Fig:3}(b). Physical origin of the resonances
in the parallel magnetic configuration, Figs.~\ref{Fig:3}(c)-(e)
and~(d)-(f), can be accounted for in a qualitatively similar way.
Finally, weak vertical lines visible in Fig.~\ref{Fig:3} for
$\omega = U$, correspond to the resonance between singly occupied
and empty OL.

\begin{figure}[t]
\includegraphics[width=0.99\columnwidth]{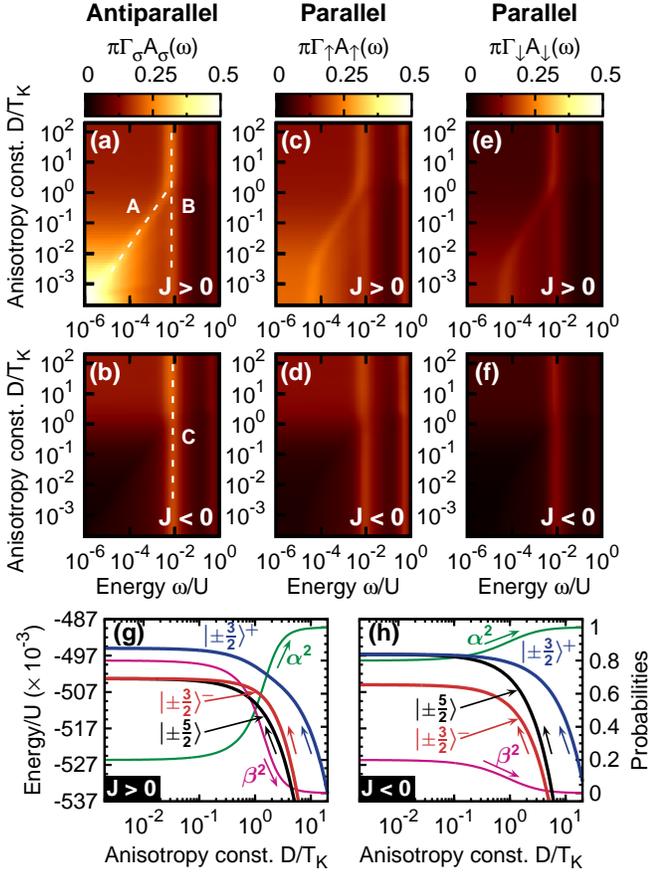}
\caption{\label{Fig:4} (Color online) Similar as in
Fig.~\ref{Fig:3}, but now the dependence of the normalized
spin-resolved spectral function on the uniaxial anisotropy
constant $D$ is presented for $|J|/T_\textrm{K}\approx2$. A
monochromatic color scheme is used here to highlight how the
position of the resonances changes with varying $D$.}
\end{figure}

As shown above, the exchange coupling of electrons in the OL and
magnetic core modifies the energy spectrum of a MQD, and hence
affects the Kondo effect. However, it has been demonstrated
experimentally~\cite{Otte_NatPhysics4/08,Parks_Science328/10,Zyazin_NanoLett10/10}
that the energy spectrum  can also be modified by changing the
anisotropy constant $D$. Variation  of the OL spectral functions
with $D$ is shown in Fig.~\ref{Fig:4} for a specific value of
$|J|/T_\textrm{K}\approx2$. Since $|J|>T_\textrm{K}$, the
zero-energy Kondo resonance is practically absent, and only side
resonances are visible. These resonances -- marked in
Figs.~\ref{Fig:4}(a)-(b) by letters A, B and C -- correspond to the
relevant resonances in Figs.~\ref{Fig:3}(a)-(b).

\begin{figure*}[t]
  \includegraphics[width=2.05\columnwidth]{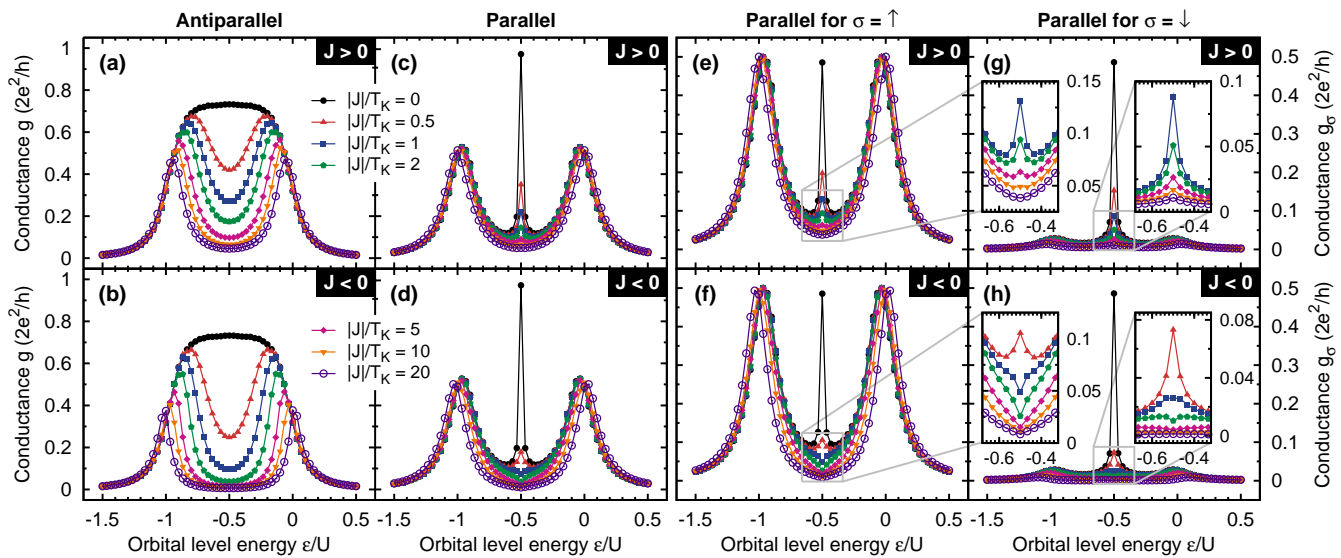}
  \caption{\label{Fig:5} (Color online) Total
linear conductance $g=\sum_\sigma g_\sigma$ in the
\emph{antiparallel} (a,b) and \emph{parallel} (c,d) magnetic
configurations, and the spin-resolved linear conductance
$g_\sigma$ in the \emph{parallel} configuration (e)-(h), presented
as a function of the OL energy $\varepsilon$ for indicated values
of the parameter $J$ in the case of the \emph{ferromagnetic}
($J>0$ -- top panel) and \emph{antiferromagnetic} ($J<0$ -- bottom
panel) exchange interaction between electrons in the OL and
magnetic core. All other parameters are as in Fig.~\ref{Fig:2}.}
\end{figure*}

Especially interesting seems to be the case of $J>0$, where two
resonances (A and B) emerge at $D\approx T_\textrm{K}$. By a
closer inspection of Fig.~\ref{Fig:4}(g) one finds that this takes
place when the condition $D\geqslant J/2$ is satisfied. However,
it should be noted that according to our definition of the Kondo
temperature, the relation $D=J/2\approx T_\textrm{K}$ is only
coincidental, and thus valid just for the current set of
parameters. Moreover, in the limit $D\gg|J|$ the energy gaps
introduced above can be estimated as
\begin{equation}\label{Eq:18}
    \left\{
        \begin{aligned}
        \Delta_1^\textrm{FM}&\approx SJ- \frac{(2S+1)^2J^2}{16(2S-1)D},
        \\
        \Delta_2^\textrm{FM}&\approx (2S-1)D + \frac{(2S+1)^2J^2}{16(2S-1)D},
        \end{aligned}
    \right.
\end{equation}
so that $\Delta_1^\textrm{FM}$ is nearly constant with respect to
$D$, whereas $\Delta_2^\textrm{FM}$ depends almost linearly on $D$.
At first glance, the latter result appears to contradict the shape
of the resonance B seen in Fig.~\ref{Fig:4}(a), so that
consideration of the energy spectrum becomes apparently
insufficient. In order to account for this disparity, we must also
take into account the explicit form of the MQD's states
contributing to the resonances, see  Eq.~(\ref{Eq:15}). For
$D>|J|/2$, we have $(\mathbb{B}^-)^2>(\mathbb{A}^-)^2$ and
$(\mathbb{A}^+)^2>(\mathbb{B}^+)^2$, and as the magnetic
anisotropy $D$ increases, at some point we get
$(\mathbb{B}^-)^2=(\mathbb{A}^+)^2\approx1$ and
$(\mathbb{A}^-)^2=(\mathbb{B}^+)^2\approx0$, Fig.~\ref{Fig:4}(g),
which basically means that
$|\!-\frac{3}{2}\rangle^-\approx|\!\uparrow\rangle_\textrm{OL}\otimes|\!-2\rangle_\textrm{core}$
and
$|\!-\frac{3}{2}\rangle^+\approx|\!\downarrow\rangle_\textrm{OL}\otimes|\!-1\rangle_\textrm{core}$.
In other words, when $D\gtrsim|J|$, the cotunneling processes
associated with transitions between the states
$|\!-\frac{5}{2}\rangle$ and $|\!-\frac{3}{2}\rangle^-$ are
allowed, while those  between the states $|\!-\frac{5}{2}\rangle$
and $|\!-\frac{3}{2}\rangle^+$ are forbidden. Consequently, the
resonance marked with a vertical dashed line in
Fig.~\ref{Fig:4}(b) actually represents two independent
resonances: B for $D\lesssim|J|$ and A for $D\gtrsim|J|$.

To complete the discussion of spectral functions, we note that the
presence and behavior of the resonance C for the antiferromagnetic
$J$-coupling ($J<0$), Fig.~\ref{Fig:4}(b), can be explained in a
way similar to that for $J>0$, with the relevant energy gaps
 for $D\gg|J|$ estimated as
\begin{equation}\label{Eq:19}
    \left\{
        \begin{aligned}
        \Delta_1^\textrm{AFM}&\approx S|J| + \frac{(2S+1)^2J^2}{16(2S-1)D},
        \\
        \Delta_2^\textrm{AFM}&\approx (2S-1)D + \frac{(2S+1)^2J^2}{8(2S-1)D}.
        \end{aligned}
    \right.
\end{equation}
However, only virtual spin-flip transitions between the states
$|\!-\frac{3}{2}\rangle^-$ and $|\!-\frac{5}{2}\rangle$, and
represented by the energy gap $\Delta_1^\textrm{AFM}$ [see
Eq.~(\ref{Eq:19})],  are then possible. On the other hand, in the
opposite limit, $D\ll|J|$, both types of spin-flip cotunneling
processes characteristic for the antiferromagnetic $J$-coupling
(as discussed earlier) can in principle operate. Both these
transitions are associated with the same energy gap given by
Eq.~(\ref{Eq:17}), which, unlike in the case of $J>0$, results in
formation of one resonance only, Fig.~\ref{Fig:4}(b).

\subsection{Conductance in the linear response regime}

From the spectral function discussed above one can determine the
spin-resolved as well as total linear conductance, shown in
Figs.~\ref{Fig:5}(a)-(g). For $|J|\ll T_\textrm{K}$, the results
well known for Kondo effect in a single-level quantum dot are
recovered.~\cite{Choi_PRL92/04,Sindel_PRB76/07,Weymann_PRB2011} In particular, for the
\emph{antiparallel} magnetic configuration an enhanced conductance
occurs in the blockade regime (single electron in the OL,
 $Q=1$), whereas in the \emph{parallel} configuration only
a sharp peak in the particle-hole symmetry point,
$\varepsilon=-U/2$, appears. More precisely, in the
\emph{antiparallel} configuration the linear conductance (measured
in the units of $2e^2/h$) is given by $g^{\rm AP} = 1-P^2$, while
in the \emph{parallel} configuration the conductance reaches unity
for $\varepsilon=-U/2$, $g^{\rm P} = 1$, with
$g^\textrm{P/AP}=\sum_\sigma g_\sigma^\textrm{P/AP}$ representing
the total linear conductance.~\cite{Weymann_PRB2011,Misiorny_NRG} Suppression of the Kondo effect for
other values of $\varepsilon$ in the latter case is a consequence
of spin splitting of the OL due to an effective exchange field
created by ferromagnetic electrodes, as it has already been
mentioned in the previous section.

When strength of the $J$-coupling increases, the Kondo effect
becomes gradually suppressed and the linear conductance in the
blockade ($Q=1$) region decreases.~\cite{Misiorny_NRG,Misiorny_JAP2011} In order to explain this
dependence, we note that in the situation under discussion
electrons are transmitted through OL that is exchange-coupled to
magnetic core, see Eqs.~(\ref{Eq:15}). As a result, cotunneling
processes responsible for the Kondo state are more complex than in
the case of $J=0$. First, the amplitude of the cotunneling
processes becomes reduced as now the electron occupying the OL is
in a superposition of the spin-up and spin-down states. Second,
the $J$-coupling creates an energy  gap between the relevant
states, which effectively suppresses the Kondo effect.

Looking more carefully at the conductance curves in
Fig.~\ref{Fig:5}, one finds some difference between the cases with
$J>0$ and $J<0$. While for $J<0$ the Kondo peak vanishes rapidly
after $|J|$ exceeds $T_\textrm{K}$, some remanent Kondo peak is
still visible for $J>0$. This disparity stems from different
properties of quantum states taking part in the formation of the
Kondo state for $J>0$ and $J<0$. First, when $|J|> T_\textrm{K}$,
the ground state energy of a singly-occupied MQD is lower for
$J<0$ than for $J>0$ [compare Figs.~\ref{Fig:3}(g) and (h)].
Second, the ground state for $J<0$ is a superposition of spin-up
and spin-down states, which is not the case  when $J>0$. As a
result, the cotunneling processes driving the Kondo effect are
more effective for $J>0$ than for $J<0$, which leads to lower
conductance for $J<0$ as compared to $J>0$.

\begin{figure}[t]
  \includegraphics[width=0.99\columnwidth]{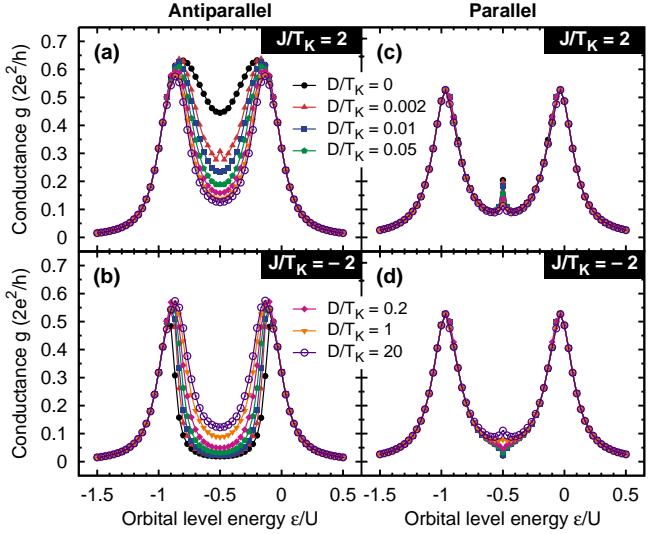}
  \caption{\label{Fig:6} (Color online) Influence of the
uniaxial magnetic anisotropy $D$ on the total linear conductance
$g=\sum_\sigma g_\sigma$ for $|J|/T_\textrm{K}\approx2$ and
the exchange $J$-coupling of either (a,c) \emph{ferromagnetic}
($J>0$) or (b,d) \emph{antiferromagnetic} ($J<0$) type. Other
parameters are the same as in Fig.~\ref{Fig:2}.
  }
\end{figure}

\begin{figure}[t]
  \includegraphics[width=0.99\columnwidth]{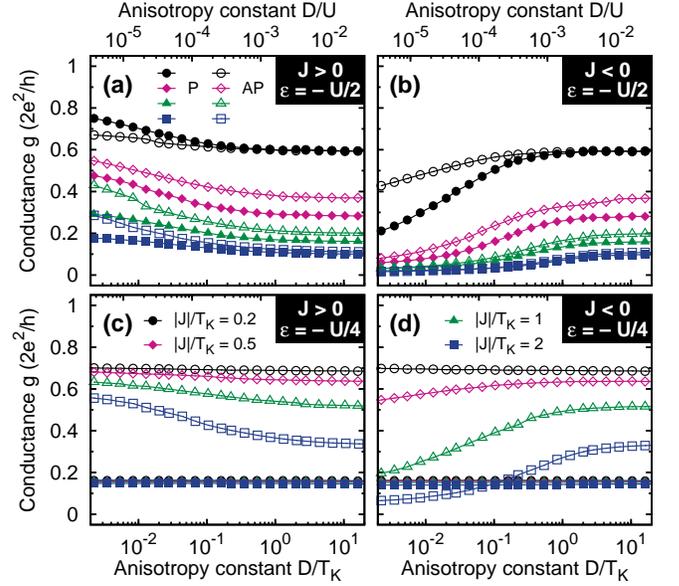}
  \caption{\label{Fig:7} (Color online) Total linear conductance $g=\sum_\sigma g_\sigma$ in
the \emph{parallel} (P -- filled points) and \emph{antiparallel}
(AP -- hollow points) magnetic configuration shown as a function
of the uniaxial anisotropy constant $D$ for indicated values of
the $J$-coupling and two representative OL energies: (a)-(b)
$\varepsilon=-U/2$ and (c)-(d) $\varepsilon=-U/4$. Left panel
corresponds to the \emph{ferromagnetic} $J$-coupling ($J>0$),
while right panel to the \emph{antiferromagnetic} one ($J<0$).
Other parameters as in Fig.~\ref{Fig:2}.
  }
\end{figure}

From the previous subsection we know that the uniaxial magnetic
anisotropy modifies electron states of a MQD, affecting thus the
Kondo effect. This is shown explicitly in Figs.~\ref{Fig:6}
and~\ref{Fig:7}. As one can note in Fig.~\ref{Fig:6}, conductance
in the parallel magnetic configuration is rather insensitive to
the anisotropy constant $D$ -- regardless of the type of exchange
coupling $J$, see the filled points in Fig.~\ref{Fig:7}(c)-(d),
and certain weak dependence on the anisotropy constant $D$ appears
then in the particle-hole symmetry point $\varepsilon=-U/2$ (Kondo
peak), see Figs.~\ref{Fig:7}(a)-(b). This dependence, however,
becomes weaker when the magnitude of exchange coupling increases.
On the other hand, the role of anisotropy is more important in the
case of antiparallel magnetic configuration (also for
$\varepsilon\neq -U/2$), especially for $|J|\gtrsim T_\textrm{K}$,
see the hollow points in Fig.~\ref{Fig:7}(c)-(d).

Variation of the conductance with the anisotropy constant $D$
depends on the sign of exchange parameter $J$. The conductance
curves for $J>0$ and $J<0$ differ significantly only for $D\ll
|J|/2$, while when $D$ exceeds $|J|/2$, the difference becomes
insignificant. Moreover, it should be noted that in the parallel
configuration the conductance for $J>0$ decreases with growing
$D$, whereas for $J<0$, one observes the opposite tendency, see
Fig.~\ref{Fig:7}. Such an overlap of the conductance curves for
ferromagnetic and antiferromagnetic $J$-coupling in the limit of
large anisotropy constant $D$ can be explained in a similar way as
above. One has to take into account that in the present situation
the difference in energy gaps between ground states for $J>0$ and
$J<0$ diminishes, and so do the energies of these states as $D$
increases.

\subsection{Tunnel magnetoresistance (TMR)}

\begin{figure}[t]
  \includegraphics[width=0.99\columnwidth]{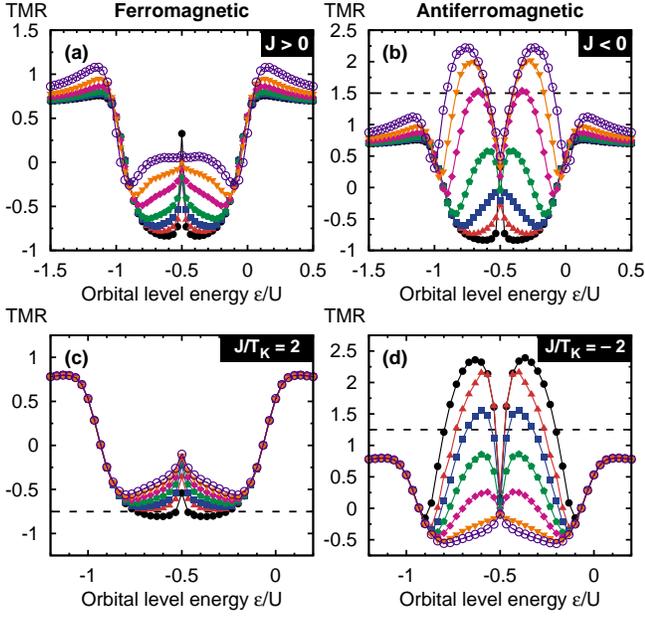}
  \caption{\label{Fig:8} (Color online) Tunnel magnetoresistance (TMR)
presented as a function of the OL energy $\varepsilon$ and the
$J$-coupling (a)-(b) and the magnetic uniaxial anisotropy $D$
(c)-(d). Specific data points and parameters used in (a)-(b)
correspond to those in Fig.~\ref{Fig:5} and, analogously, in
(c)-(d) to Fig.~\ref{Fig:6}.
Dashed lines are introduced in order to facilitate comparison
between TMR scales of adjacent plots for $J>0$ and $J<0$.
  }
\end{figure}

A quantity that describes difference  between transport properties
in the \emph{parallel} (P) and \emph{antiparallel} (AP) magnetic
configurations is the \emph{tunnel magnetoresistance} (TMR),
defined here as~\cite{Julliere_PLA54/75}
\begin{equation}\label{Eq:20}
    \textrm{TMR}=\frac{g^\textrm{P}-g^\textrm{AP}}{g^\textrm{AP}}.
\end{equation}
Using the conductance data analyzed in the previous subsection, we
consider now how TMR depends on the key parameters of the model,
i.e. on the $J$-coupling, Figs.~\ref{Fig:8}(a)-(b) and the magnetic
anisotropy $D$, Figs.~\ref{Fig:8}(c)-(d).

Let us first discuss the behavior of TMR as a function of the
exchange coupling constant $J$, Figs.~\ref{Fig:8}(a)-(b). Since for
$|J|\ll T_\textrm{K}$ the conductance in the Kondo regime is generally
larger in the antiparallel magnetic configuration than in the
parallel one, regardless of the type of the $J$-coupling, the
corresponding TMR is negative in almost the entire Coulomb
blockade region. The only exception occurs around the
particle-hole symmetry point
$\varepsilon=-U/2$.~\cite{Misiorny_NRG} This behavior follows from
the suppression of the Kondo effect in the parallel configuration due to the exchange field,
except for the particle-hole symmetric point. However, as $|J|$
becomes larger than  $T_\textrm{K}$, the Kondo peak becomes
suppressed also in the antiparallel configuration and positive TMR
may be observed in the blockade regime as well. Moreover,
suppression of the conductance $g^{\rm AP}$ for the antiparallel
alignment is more evident in the case of \emph{antiferromagnetic}
($J<0$) coupling, Fig.~\ref{Fig:8}(b), and the corresponding TMR
is therefore significantly larger than for the
\emph{ferromagnetic} coupling ($J>0$). Another observation for
$J<0$ is that when $|J|\gtrsim T_\textrm{K}$, two distinctive
local maxima develop in the Coulomb blockade regime. Their
positions depend on $J$ and are symmetrical with respect to
$\varepsilon=-U/2$. In addition, for $J<0$, the TMR considerably
surpasses the relevant Julliere's value,
$2P^2/(1-P^2)$,~\cite{Julliere_PLA54/75} which for the present
parameters yields $2/3$. Nonetheless, in the regions corresponding
to empty or doubly occupied OL, one always observes $g^{\rm P} >
g^{\rm AP}$ with TMR approaching the Julliere's value.~\cite{Weymann_PRB2011}

Consider now variation of TMR with the magnetic anisotropy $D$,
Figs.~\ref{Fig:8}(c)-(d). First of all, when the OL is either
empty or occupied by two electrons, TMR remains insensitive to any
changes in the anisotropy constant $D$. The same cannot be said
about the region of $\varepsilon$ corresponding to single
occupation of the OL, where large variation of TMR appears
especially for $J<0$, see Fig.~\ref{Fig:8}(d). From
Fig.~\ref{Fig:8}(d) follows that the smaller the magnetic
anisotropy, the larger TMR. Positions of the two local maxima in
TMR, however, are now rather independent of $D$. Furthermore, for
$D\ll |J|/2$, TMR stays positive. On the contrary, for the
ferromagnetic $J$-coupling ($J>0$), Fig.~\ref{Fig:8}(c), the TMR
is negative in the considered range of $\varepsilon$ and is only
weakly modified upon changing $D$.

\subsection{The restoring effect of magnetic field}

\begin{figure}[t]
  \includegraphics[width=0.99\columnwidth]{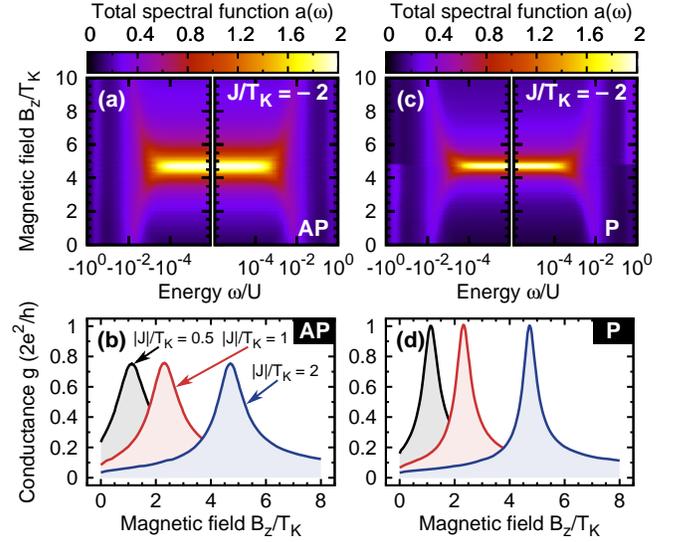}
  \caption{\label{Fig:9} (Color online)  Restoration of the Kondo peak in
  the total spectral function
$a(\omega)=\pi\sum_\sigma\Gamma_\sigma A_\sigma(\omega)$ (a,c),
and total conductance $g=\sum_\sigma g_\sigma$ (b,d) by an
external magnetic field $B_z$, shown for indicated values of the
\emph{antiferromagnetic} $J$-coupling ($J<0$) and for
$\varepsilon=-U/2$. Left panel corresponds to the
\emph{antiparallel} (AP) magnetic configuration, while right panel
to the \emph{parallel} (P) one. The other  parameters as in
Fig.~\ref{Fig:2}.
  }
\end{figure}

In the light of the foregoing discussion, we know that the Kondo
effect is suppressed by exchange field generated by ferromagnetic
electrodes as well as by the exchange coupling of the OL to
magnetic core. Very recently, it was shown both experimentally and theoretically
that one can compensate for the exchange-induced splitting of the orbital
level by fine-tuning an external magnetic field, restoring thus the universal features of the Kondo effect.~\cite{Gaass_2011}
Now we will thus consider the possibility of restoring
the Kondo effect by applying a compensating external magnetic field, $B_{\rm c}$, oriented along the
MQD's easy axis, Fig.~\ref{Fig:9}. The interesting observation
worth mentioning is that the restoration can take place not only
for the parallel magnetic configuration, Figs.~\ref{Fig:9}(c)-(d),
but also for the antiparallel one, Figs.~\ref{Fig:9}(a)-(b).
Furthermore, we would like to emphasize that the considered effect
occurs only for the \emph{antiferromagnetic} ($J<0$) coupling
between an electron in the OL and the MQD's magnetic core. From
Figs.~\ref{Fig:9}(c)-(d) follows that for the parallel magnetic
configuration the full unitary Kondo resonance (with the maximum
value of conductance $g=1$) is restored, whereas in the
antiparallel configuration the peak of height $1-P^2$ is retrieved
owing to the magnetic field.

\begin{figure}[t]
  \includegraphics[width=0.99\columnwidth]{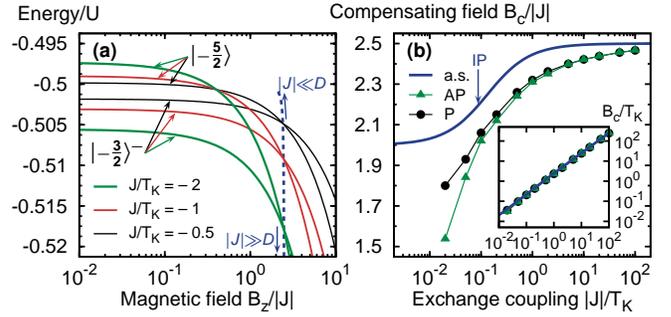}
  \caption{\label{Fig:10} (Color online)  (a) Energies of ground
states $|\!-\tfrac{5}{2}\rangle$ and $|\!-\tfrac{3}{2}\rangle^-$
shown as functions of an external magnetic field $B_z$ for
indicated  values of the exchange coupling parameter $J$ and
for $D/U\approx1.7\cdot10^{-4}$ ($D/T_\textrm{K}=0.1$) and $\varepsilon=-U/2$. Dashed line
corresponds to the analytical solution given by Eq.~(\ref{Eq:21}).
(b) Dependence of the compensating field $B_\textrm{c}$ on the value
of the $J$-coupling. Points represent numerical results obtained
for the \emph{antiparallel} (AP) and \emph{parallel} (P) magnetic
configuration, while the bold line delineates the analytical
solution (a.s.), Eq.~(\ref{Eq:21}). IP indicates the inflexion point of
the curve $B_\textrm{c}(|J|)$. Inset: the compensating field $B_{\rm c}$
normalized to $T_\textrm{K}$, shown in the logarithmic scale.
Other parameters as in Fig.~\ref{Fig:2}.}
\end{figure}

From the experimental point of view, it would be useful to know,
at least roughly, how the magnitude of the compensating field
$B_\textrm{c}$ depends on the system's parameters. For this
purpose, it is essential to know physical mechanism responsible
for the restoration of the Kondo effect. Thus, in order to gain
some deeper understanding of the problem once again we  employ the
model of a free-standing MQD. It appears that the restoration of
the Kondo effect becomes possible always when the magnetic field
$B_z$ brings the ground state of the spin multiplet $S_t=S-1/2$
into resonance with the ground state of the multiplet $S_t=S+1/2$.
Depending on the direction of the field, the resonance takes place
either between the states
$|\!-\frac{3}{2}\rangle^-\leftrightarrow|\!-\frac{5}{2}\rangle$ or
$|\!+\frac{3}{2}\rangle^-\leftrightarrow|\!+\frac{5}{2}\rangle$.
Moreover, the possibility of such degeneracy is straightforwardly
granted by the fact that these states are characterized by
different numbers $S_z^t$ (different $z$th component of the MQD's
total spin). Accordingly, the Zeeman contributions to these states
are different. Because the $J$-coupling is of the
\emph{antiferromagnetic} type, the state with greater $|S_z^t|$
has then larger energy in the absence of magnetic field ($B_z=0$).
Thus, if $B_z>0$, the energy of the state $|\!-\frac{5}{2}\rangle$
decreases faster with increasing the field than the energy of
$|\!-\frac{3}{2}\rangle^-$, so the degeneracy of these states
occurs at a certain value of magnetic field, $B=B_\textrm{c}$, which can be then
determined from the condition
$\Delta_1^\textrm{AFM}(B_\textrm{c})=0$, Fig.~\ref{Fig:10}(a).

Taking into account the resonance condition introduced in the
previous paragraph, we find the general expression describing
dependence of the compensating field $B_\textrm{c}$  on the
$J$-coupling and magnetic anisotropy $D$ in the form
\begin{align}\label{Eq:21}
    B_\textrm{c}&=\frac{2S+1}{4}|J|-\frac{2S-1}{2}D
    \nonumber\\
    &+\sqrt{\frac{(2S+1)^2}{16}J^2+\frac{(2S-1)^2}{4}D(D+|J|)}\,.
\end{align}
The comparison between the analytical solution and numerically derived
values of the compensating field $B_\textrm{c}$ for different
magnetic configurations of the system is presented in
Fig.~\ref{Fig:10}(b). It is clearly visible that the numerical
values of $B_\textrm{c}$ generally follow the trend of the
approximate analytical curve, Eq.~(\ref{Eq:21}), and substantial
discrepancies arise only for small values of $|J|$, as one might
expect. The slightly smaller value of the compensating field in the
situation of a MQD attached to magnetic electrodes can be
attributed to renormalization of MQD's energy levels due to the
strong OL-electrodes coupling, which results in diminishing energy
gaps between the states participating in formation of the Kondo
effect. It should be here emphasized that this difference can be seen
only due to specific normalization of the compensating field $B_{\rm c}$ with respect to $|J|$.
Otherwise, when presented in the logarithmic scale,
the curves corresponding to numerical and analytical
solutions follow the same trend, see the inset in Fig.~\ref{Fig:10}(b).

Finally, since the form of Eq.~(\ref{Eq:21}) is exactly the same
as for $\Delta_1^\textrm{AFM}$, Eq.~(\ref{Eq:B2}), the asymptotic
values of $B_\textrm{c}$ are immediately obtained as:
$B_\textrm{c}\approx(2S+1)|J|/2$ for $|J|\gg D$, and
$B_\textrm{c}\approx S|J|$ for $|J|\ll D$. Additionally, it might
be helpful to know how the change of magnetic anisotropy $D$
influences the  analytical solution $B_\textrm{c}(|J|)$, bold line
in Fig.~\ref{Fig:10}(b). For this purpose, we calculate the
inflexion point position (IPP) of the compensating magnetic field
curve $B_\textrm{c}(|J|)$,
\begin{equation}\label{Eq:22}
    \textrm{IPP}=
    \frac{2(2S-1)}{(2S+1)^2}\Big[1-(2S)^{2/3}\Big]\Big[1+(2S)^{1/3}\Big]D.
\end{equation}
It turns out that IPP depends linearly on $D$, with the
proportionality constant being a complex function of the MQD's
core spin number $S$. Consequently, modification of the magnetic
anisotropy constant $D$ does not affect the general shape of the
compensating field curve $B_\textrm{c}(|J|)$, but it only leads to
shifting of the curve's inflexion point.

\section{Summary and conclusions}

In this paper we have investigated transport properties in the
Kondo regime of a class of systems exhibiting uniaxial magnetic
anisotropy. The model assumed includes one orbital level through
which electrons can tunnel and which is additionally
exchange-coupled to a magnetic moment. The model  can describe a
single-level quantum dot in which electrons are exchange-coupled
to an embedded magnetic impurity. It can also be used to describe
transport through magnetic atoms and molecules.

Using the numerical renormalization group method we have
calculated spectral density of the OL level and linear conductance
of the system. The key new feature of
transport characteristics is the suppression of the Kondo effect
by exchange coupling to magnetic core. Independently of the sign
of the coupling parameter $J$,  suppression takes place for both
ferromagnetic as well as antiferromagnetic coupling. In the limit
of small $J$, $|J|\ll T_\textrm{K}$, we find the Kondo resonance
characteristic of single-level dots. This resonance is suppressed
with increasing $J$ but then some side resonances appear in the
spectral density. It is worthy of note that the suppression of the
Kondo peak in the case considered appears gradually with
increasing  $J$, contrary to the case of a dot exchange-coupled to
electron reservoir of continuous density of states, where the
suppression is associated with a quantum phase transition.~\cite{Weymann_PRB81/10}

We have also shown that the suppressed Kondo resonance in spectral
function and in transport characteristics can be restored by
application of an external magnetic field. This restoration is
however complete only in the case of antiferromagnetic exchange
coupling ($J<0$) for both the parallel and antiparallel magnetic
configurations.


\begin{acknowledgments}

This work was supported by Polish Ministry of Science and Higher
Education through a research project in years 2010-2013 and
a `Iuventus Plus' research project for years 2010-2011. I.W.
also acknowledges support from the Polish Ministry of Science and
Higher Education through a research project in years 2008-2011 and
the Alexander von Humboldt Foundation.
We acknowledge usage of the Budapest NRG
code~\cite{Legeza_DMNRGmanual} in numerical calculations,
which has been also employed
in Refs. [\onlinecite{Misiorny_NRG}] and [\onlinecite{Misiorny_JAP2011}].

\end{acknowledgments}

\appendix
\section{\label{APP:A}Calculation of the spectral function}

During the course of NRG iterative calculation, the spectral
function $A_\sigma(\omega)$ is obtained as a set of the Dirac
delta functions $\delta(\omega-\omega_n)$ at frequencies
$\omega_n$, which next have to be broaden in order to acquire a
continuous spectrum. Because of logarithmic discretization of
conduction band, it is convenient to collect the delta peaks in
logarithmic bins and broaden them using a logarithmic Gaussian
distribution, with the broadening parameter typically given by
$\log(\Lambda)$.~\cite{Sakai_JPhysSocJapan58/89,Costi_JPhysCondensMatter6/94,Zitko_PRB79/09}
It turns out, however, that due to logarithmic discretization of
the band and truncation during the NRG run, the broadened spectral
function may exhibit some artifacts, such as an oscillatory
behavior for energies smaller than the Kondo temperature. One of the tricks
to overcome these problems is
the so-called {\it self-energy trick}.~\cite{Zitko_PRB79/09,Bulla_JPhysCM10/98}

The essential idea of the {\it self-energy trick} relies on finding the spectral function
by constructing the full self-energy $\Sigma_\sigma(\omega)$ of the
system.~\cite{Bulla_JPhysCM10/98} The self-energy can be expressed
as a ratio of two spectral functions. Although each spectral
function displays similar problems related with discretization, by
calculating their ratio one obtains a smooth function where most
of the artifacts are suppressed. Having found the self-energy, one
can then calculate the retarded Green's function $\langle\langle
c_\sigma^{} | c_\sigma^\dag \rangle\rangle_\omega^\textrm{r}$ of
the OL,  Eq.~(\ref{Eq:12}), using the equation of motion
method~\cite{Jauho_book}
\begin{equation}\label{Eq:A2}
    \langle\langle c_\sigma^{} | c_\sigma^\dag \rangle\rangle_\omega^\textrm{r}=
    \frac{1}{\omega-\varepsilon_\sigma-\Sigma_\sigma^\textrm{r}(\omega)},
\end{equation}
where $\varepsilon_\sigma = \varepsilon +
\frac{1}{2}B_z(\delta_{\sigma\uparrow}-\delta_{\sigma\downarrow})$.
The total self-energy consists of three terms,
\begin{equation}\label{Eq:A3}
    \Sigma_\sigma^\textrm{r}(\omega)=\Sigma_\sigma^{U\textrm{r}}(\omega)+\Sigma_\sigma^{J\textrm{r}}(\omega)+
    \Delta_\sigma^\textrm{r}(\omega),
\end{equation}
which represent contributions stemming from: the Coulomb
interaction, $\Sigma_\sigma^{U\textrm{r}}(\omega)$, the exchange
interaction between an electron spin in the OL and the core spin
of the MQD, $\Sigma_\sigma^{J\textrm{r}}(\omega)$, and the
tunneling coupling of the OL to electrodes,
$\Delta_\sigma^\textrm{r}(\omega)$. The explicit forms of the
above self-energy contributions are given by
\begin{equation}\label{Eq:A4}
    \Sigma_\sigma^{U\textrm{r}}(\omega)=
    U\frac{\langle\langle n_{\overline{\sigma}}^{}c_\sigma^{}| c_\sigma^\dag \rangle\rangle_\omega^\textrm{r}}
    {\langle\langle c_\sigma^{} | c_\sigma^\dag \rangle\rangle_\omega^\textrm{r}},
\end{equation}
\begin{multline}\label{Eq:A5}
    \Sigma_\sigma^{J\textrm{r}}(\omega)=-\frac{J}{2}\Bigg[
    \delta_{\sigma\uparrow}\frac{\langle\langle c_{\overline{\sigma}}^{} S_- | c_\sigma^\dag \rangle\rangle_\omega^\textrm{r}}{\langle\langle c_\sigma^{} | c_\sigma^\dag \rangle\rangle_\omega^\textrm{r}}
    +\delta_{\sigma\downarrow}\frac{\langle\langle c_{\overline{\sigma}}^{} S_+ | c_\sigma^\dag \rangle\rangle_\omega^\textrm{r}}{\langle\langle c_\sigma^{} | c_\sigma^\dag \rangle\rangle_\omega^\textrm{r}}
    \\
    +\big(\delta_{\sigma \uparrow} - \delta_{\sigma \downarrow}\big)
    \frac{\langle\langle c_{\sigma}^{} S_z | c_\sigma^\dag \rangle\rangle_\omega^\textrm{r}}{\langle\langle c_\sigma^{} | c_\sigma^\dag \rangle\rangle_\omega^\textrm{r}}\Bigg],
\end{multline}
\begin{equation}\label{Eq:A6}
    \Delta_\sigma^{\textrm{r}}(\omega)=\frac{\Gamma_\sigma}{\pi} \Bigg[
    \ln\Big|\frac{W+\omega}{W-\omega}\Big|- i\pi
    \Bigg].
\end{equation}
Finally, the {\it improved} spectral function of the orbital level
is given by
\begin{equation}\label{Eq:A7}
\hspace*{-2pt}
    A_\sigma(\omega)=-\frac{1}{\pi}\!\cdot\!\frac{
    \textrm{Im}\Sigma_\sigma^{\textrm{r}}(\omega)
    }
    {
    \big[\omega-\varepsilon_\sigma-\textrm{Re}\Sigma_\sigma^{\textrm{r}}(\omega)\big]^2
    \!+\!\big[\textrm{Im}\Sigma_\sigma^{\textrm{r}}(\omega)\big]^2
    }.
\end{equation}

\section{\label{APP:B}Energy gaps}

In the case of the \emph{ferromagnetic} (FM) $J$-coupling ($J>0$),
exact analytical expressions for the energy gaps
$\Delta_1^\textrm{FM}$  and $\Delta_2^\textrm{FM}$ can be derived as
the energy difference between appropriate MQD's states:
$|\!\pm\frac{5}{2}\rangle$ and $|\!\pm\frac{3}{2}\rangle^-$ for
$\Delta_1^\textrm{FM}$, and  $|\!\pm\frac{5}{2}\rangle$ and
$|\!\pm\frac{3}{2}\rangle^+$ for
$\Delta_2^\textrm{FM}$,~\cite{Note1} so that
\begin{align}\label{Eq:B1}
    \Delta_{1(2)}^\textrm{FM}&=\frac{2S+1}{4}|J|+\frac{2S-1}{2}D
    \nonumber\\
    &\mp\sqrt{\frac{(2S+1)^2}{16}J^2+\frac{(2S-1)^2}{4}D(D-|J|)}.
\end{align}
The energy gaps for \emph{antiferromagnetic} (AFM) $J$-coupling ($J<0$)
can be found in a similar way. The energy difference between the
MQD's states $|\!\pm\frac{3}{2}\rangle^-$ and
$|\!\pm\frac{5}{2}\rangle$ is then given by
\begin{align}\label{Eq:B2}
    \Delta_{1}^\textrm{AFM}&=\frac{2S+1}{4}|J|-\frac{2S-1}{2}D
    \nonumber\\
    &+\sqrt{\frac{(2S+1)^2}{16}J^2+\frac{(2S-1)^2}{4}D(D+|J|)},
\end{align}
whereas the formula for the gap between the states
$|\!\pm\frac{3}{2}\rangle^-$ and $|\!\pm\frac{3}{2}\rangle^+$
takes the form
\begin{equation}\label{Eq:B3}
    \Delta_{2}^\textrm{AFM}=2\sqrt{\frac{(2S+1)^2}{16}J^2+\frac{(2S-1)^2}{4}D(D+|J|)}.
\end{equation}
%



%

\end{document}